\documentclass[aps,
12pt,showpacs,showkeys
,twoside]{revtex4}

\usepackage{graphicx}
\usepackage{color}
\usepackage{epsfig}
\usepackage{amsmath}
\usepackage{amstext}
\usepackage{amssymb}
\usepackage{bm}
\usepackage[english]{babel}
\usepackage[latin2]{inputenc}
\DeclareGraphicsExtensions{.eps,.ps}
\begin{document}
\author{Antoni W{\'o}jcik }
\email[Electronic address: ]{antwoj@amu.edu.pl}

\author{ Ravindra W. Chhajlany }
\email{ravi@amu.edu.pl}

\affiliation{Faculty of Physics, Adam Mickiewicz University,
 \\ Umultowska 85,  61-614 Pozna{\'n}, Poland}
 \date{\today}

\title{Illusion of quantum speed-up}

\begin{abstract}
  Quantum computers are believed to surpass classical ones. Moreover,
  it is claimed that this belief reaches the level of a mathematically
  proven fact within the so-called oracle model of computation. Here
  we impair the whole class of the so-called rigorist proofs of
  quantum speed-up obtained within this  model.
\end{abstract}




\maketitle

Among the reasons underlying recent interest in quantum information
processing is a ``reasonable hope'' \cite{ben00} that quantum
computers could speed up solving certain problems. This belief is
supported by many results within the oracle model of computation,
which are usually interpreted as rigorist proofs of quantum over
classical computation superiority.  We show, however, that the
generally accepted method of comparing quantum and classical oracles,
which is a cornerstone of these proofs, is inconsistent and that the
quantum speed-up can disappear when the above-mentioned inconsistency
is removed.  Let us take as an example the Bernstein-Vazirani problem
(BVP)\cite{bern97, ter98}: a $n$-bit string $\vec{k}$ is embodied in
an oracle and the goal is to identify $\vec{k}$. The classical oracle
$O_{S}$ (in the so-called standard form) transforms a ($n+1$)-bit
input string $\vec{x}= (x_{0}, x_{1}, \ldots, x_{n})$ into an output
string according to the following rules
\begin{gather}
\begin{array}{l}
 x_{0} \xrightarrow{O_{S}} x_{0} \oplus \vec{k} \cdot \vec{x}
\\
x_{j} \xrightarrow{O_{S}} x_{j}  \; (j=1,2, \ldots n)
\end{array}
\label{Rgpa}
\end{gather}
where $\vec{k} \cdot \vec{x} = k_{1}x_{1}\oplus \ldots \oplus
k_{n}x_{n}$ and $\oplus$ denotes addition modulo 2. 

On the other hand the quantum oracle $U_{S}$ is given in the form of a
unitary operator acting on a string of qubits instead of bits.
Although the algorithms which call different oracles should not be
compared, it is generally accepted to compare the ``corresponding''
classical and quantum oracles. To establish such a correspondence, a
computational basis $|\vec{z} \rangle = |z_{0}\rangle |z_{1}\rangle
\ldots |z_{n}\rangle $ is defined, {\it i.e.} for each qubit two
orthonormal states are chosen and labeled $|z_{j}\rangle $ ($z=0,1$).
Note that the choice of these states is arbitrary and can be made
independently for each qubit. Having defined computational states,
$U_{S}$ can now be identified by giving its action on these states
only
\begin{gather}
\begin{array}{l}
|z_{0}\rangle \xrightarrow{U_{S}} |z_{0} \oplus \vec{k} \cdot
 \vec{z}\rangle \\
 |z_{j}\rangle \xrightarrow{U_{S}} |z_{j}\rangle, \; (j=1,2, \ldots n) .
\end{array}
\label{Rhpa}
\end{gather}
The correspondence desired is based just on a formal identity of the
transformation rules defining $U_{S}$ and $O_{S}$ (see
Eqs.(\ref{Rgpa}) and (\ref{Rhpa})). Comparison of two algorithms - the
optimal classical algorithm, which needs $n$ queries to $O_{S}$, with
the famous Bernstein-Vazirani quantum algorithm \cite{cleve98} solving
the problem with just a single query to $U_{S}$, provides the proof of
the quantum speed-up in BVP.  Our criticism of this proof starts with
noting that $O_{S}$ is not a unique oracle that can be considered as a
classical counterpart (CCP) of $U_{S}$.  Imagine, {\it e.g.}, that
Alice, Bob and Steven are asked to prepare quantum oracles
corresponding to three classical oracles $O_{A}, O_{B}$ and $O_{S}$,
respectively. $O_{A}$ and $O_{B}$ are defined in the following way
\begin{gather}
\begin{array}{l}
x_{0} \xrightarrow{O_{A}} x_{0} \\
x_{j} \xrightarrow{O_{A}} x_{j}+ k_{j} x_{0} \; (j=1,2, \ldots, n),
\end{array}
\label{Ripa}
\end{gather}
\begin{gather}
\begin{array}{l}
x_{n} \xrightarrow{O_{B}} x_{n} + k_{n} x_{0} \\
x_{j} \xrightarrow{O_{B}} x_{j} \; (j=0,1, \ldots, n-1).
\end{array}
\label{Rjpa}
\end{gather}
Note that BVP can be solved with a single query to $O_{A}$, and cannot
be solved at all with the use of $O_{B}$.  What can come as a
surprise, is that all three parties can prepare the same quantum
oracle $U_{A}=U_{B}=U_{S}$.  This can happen because each party can
define the computational basis in a different way.  To see this let us
denote by $|\uparrow \rangle $ and $|\downarrow \rangle $ two
arbitrary orthogonal states which span each single qubit Hilbert
space. Steven chooses these states as his computational states, {\it
  i.e.}
\begin{gather}
\begin{array}{l}
|0_{j}\rangle = |\uparrow  \rangle \\
|1_{j}\rangle = |\downarrow  \rangle ,
\end{array}
\label{Rkpa}
\end{gather}
for $j=0, 1, \ldots n$. On the other hand Alice's choice is (for
$j=0,1, \ldots, n$)
\begin{gather}
\begin{array}{l}
|0_{j}\rangle = 2^{-1/2}(|\uparrow  \rangle +
 |\downarrow  \rangle) \\
|1_{j}\rangle = 2^{-1/2}(|\uparrow  \rangle - |\downarrow  \rangle ).
\end{array}
\label{Rmpa}
\end{gather}
It follows that her quantum oracle $U_{A}$ although defined by
correspondence with the rules given by Eq.(\ref{Ripa}) is identical to
Steven's oracle $U_{S}$. Bob takes advantage of the arbitrariness in
defining each single qubit computational basis states by choosing 
\begin{gather}
\begin{array}{l}
|0_{j}\rangle = |\uparrow  \rangle \\
|1_{j}\rangle = |\downarrow  \rangle
\end{array}
\label{Rnpa}
\end{gather}
for $j=1, \ldots n-1$ and 
\begin{gather}
\begin{array}{l}
|0_{j}\rangle = 2^{-1/2}(|\uparrow  \rangle +
 |\downarrow  \rangle) \\
|1_{j}\rangle = 2^{-1/2}(|\uparrow  \rangle - |\downarrow  \rangle ).
\end{array}
\label{Ropa}
\end{gather}
for $j=0,n$. Similarly to Alice's case, the oracle of Bob will again
be $U_{B}=U_{S}$.  Obviously, there is no reason to favour any
particular choice of computational basis. Thus both $O_{A}$ and
$O_{B}$ should be considered as CCP of $U_{S}$ as well as $O_{S}$.
Now, the basic question arises: to which of its CCPs - $O_{A}$,
$O_{B}$ or $O_{S} $ should $U_{S}$ be compared?  Quantum speed-up
obtained by comparing $U_{S}$ to $O_{S}$ disappears when the latter is
replaced by $O_{A}$, whereas it approaches infinity in the case of
$O_{B}$. This ambiguity presents a serious challenge, which can be
approached in two ways. The first approach simply enforces the earlier
mentioned statement that different oracles ({\it i.e.}  quantum and
classical) should not be compared. On the other hand, reliable
estimation of quantum speed-up seems to be still possible, provided
that the optimal CCP of quantum oracle is found and used for
comparison. For example, in the case of BVP the quantum oracle $U_{S}$
should not be compared with the classical oracle $O_{S}$ but with
$O_{A}$. Both $U_{S}$ and $O_{A}$ allow a single query solution of the
problem. It follows that the claimed quantum speed-up in BVP is just
an artefact of the non-optimal choice of computational basis.  The
above conclusion, although illustrated by the BVP example, is in fact
general. In the light of this reasoning all the so-called rigorist
proofs of quantum speed-up obtained previously within the oracle model
of computation must be reviewed. This calls for the solution of the
nontrivial problem of finding optimal CCP of a given unitary oracle if
a reliable comparison of oracles is to be made.

\acknowledgments
  
A. W. would like to thank the State Commision for Scientific Research
for financial support under grant no. 0 T00A 003 23.

\end{document}